\documentclass[%
 prb,
 amsmath,amssymb,
 reprint,%
 longbibliography,
 hidelinks,
]{revtex4-1}

\usepackage{graphicx}
\usepackage{dcolumn}
\usepackage{bm}

\usepackage{multirow}
\usepackage{textcomp}
\graphicspath{{fig/}}
\usepackage{hyperref}
\usepackage{enumitem}

\begin{document}


\title{Generalized Fit for Asymptotic Predictions of Heavy-Tail Experimental Transients}
\author{Jiajun Luo}
\affiliation{
Department of Electrical Engineering and Computer Science, Northwestern University, Evanston, IL 60208, USA
}
\author{M. Grayson}
\email[Corresponding author: ]{mgrayson@eecs.northwestern.edu}
\affiliation{
Department of Electrical Engineering and Computer Science, Northwestern University, Evanston, IL 60208, USA
}

\date{\today}

\begin{abstract}
	Transient responses in disordered systems typically show a heavy-tail relaxation behavior: the decay time constant increases as time increases, revealing a spectral distribution of time constants. The asymptotic value of such transients is notoriously difficult to experimentally measure due to the increasing decay time-scale.  
	However, if the heavy-tail transient is plotted versus log-time, a reduced set of data around the inflection point of such a plot is sufficient for an accurate fit. 
	From a derivative plot in log-time, the peak height, position, line width, and, most importantly, skewness are all that is needed to accurately predict the asymptotic value of various heavy-tail decay models to within less than a percent.  This curve fitting strategy reduces by orders of magnitude the amount of experimental data required, and clearly identifies a threshold below which the amount of data is insufficient to distinguish various models.
	The skew normal spectral fit and dispersive diffusion transient fit are proposed as four-parameter fits, with the latter including the stretched exponential as a limiting case.
	The line fit and asymptotic prediction are demonstrated using experimental transient responses in previously published amorphous silicon and amorphous InGaZnO data.
	
\end{abstract}


\maketitle

\section{Introduction}
Transient responses in disordered physical systems frequently exhibit a range of relaxation time scales.  When rapid decay scales govern the initial response but ever slower decay scales govern the longer time response, such relaxations are called heavy-tail transients.  For example, dielectric relaxation in cold glasses well below the glass transition temperature in the presence of a static electric field represents the first system where this behavior was observed by Kohlrausch in 1854. \cite{Kohlrausch1854} The same behavior was rediscovered by Williams and Watts over a century later \cite{Williams1970}, and the legacy of glassy systems has been inherited by the soft-matter community where reviews of the empirical behavior have been written, \cite{Phillips1996} though there is a tendency to rely on stretched exponentials to fit the data, \cite{Hall1997, Fukao2000, Bohmer1993, Lee2008} without considering other heavy-tail candidate functions. \cite{Ngai1981, Angell2000}
In addition, biological systems such as stress-relaxation in bone and cartilage, \cite{June2013} seismic systems such as earthquake magnitude and frequency, \cite{Klein2008} and complex stochastic systems such as stock market survival probability \cite{Raberto2002} all exhibit heavy-tail transients.

This manuscript focuses on the heavy-tail transients that occur in technologically relevant electronic materials, such as the photoconductivity response in amorphous semiconductors, \cite{SEbook} as well as in systems that one might not initially think of as disordered, such as the field-induced conductivity transients in two-dimensional materials on imperfect substrates \cite{Wu2015,Late2012} and photoluminescence of type-II superlattices suffering from interface diffusion. \cite{Olson2012} In all the above cases, the heavy-tail transients occur when switching from one steady-state condition to another.
But across different scientific communities, there are advances in knowledge that are accepted in one community that are unknown in another, so a clear unified summary of the heavy-tail phenomenology is needed.  Furthermore, new developments, such as the four-parameter skewed-fit proposed here, can have broad impact across many fields.

Heavy-tail transients can always be mathematically decomposed into an amplitude spectrum of exponential decays spanning a range of time constants, whereby the shape of the time constant spectrum serves as a fingerprint of the physical mechanism behind the heavy-tail decay.
For example, this spectrum of time constants can be discrete in crystals with multiple species of ionized dopants, whereby a different relaxation time constant is associated with each defect level, forming the basis of photoinduced current transient spectroscopy (PICTS) experiments.\cite{Eiche1992a, Mathew2003}
In disordered systems, on the other hand, one expects a continuum of time constants. Amorphous semiconductors also exhibit heavy-tail transients, \cite{Lee2010, Redfield1989, Luo2013} and although some authors have similarly modeled these with a handful of discrete relaxation time constants,\cite{Lee2010, Studenikin1998} the justification for such an interpretation is unclear due to the structural randomness which seems to preclude populations of nominally identical defects. Instead, transients in amorphous systems typically assume a continuum of time constants which are statistically distributed within a broad range.\cite{Luo2013, Nagase1999, Kim2012b}

Strong disorder potentials with a continuum of relaxation time constants can also be found in crystalline semiconductor systems.  For example, two-dimensional materials are unable to screen a nearby disorder potential due to their atomic thickness and their low-dimensional density of states. The proximity of the disorder potential caused by the substrate interface roughness and remote ionized impurities \cite{Wu2015} results in a strong disorder potential, leading to a continuum of relaxation time constants.
A disorder-induced range of time constants is also expected in photoluminescence decay in type II superlattices since the forward diffusion of the column V atoms at the superlattice interfaces creates a disorder potential that separates localized electrons and holes at random distances. \cite{Steenbergen2011, Connelly2016} Thus the characterization of heavy-tail transients has broad relevance in various modern electronic materials research including some crystalline systems.

Several simple analytical functions have been proposed to describe such heavy-tail transients and/or the resulting time constant spectrum with just a few parameters.
The most well-known example is the stretched exponential function, also known as the Kohlrausch-Williams-Watts (KWW) function, \cite{Kohlrausch1854, Williams1970} which adds a stretching exponent $\beta$ to the simple exponential function
\begin{equation}
	\label{eq:SE_PPC_intro}
	f_{SE}(t) = f_{\infty} + A \mathrm{e}^{-(t/\tau_0)^{\beta}} .
\end{equation}
The stretched exponential provides an excellent empirical fit to certain heavy-tail experimental transients observed in amorphous semiconductors and polymers over a wide range of time scales. \cite{Luo2013,Kakalios1987, Lee2009b}
Alternatively, the less well-known algebraic decay or inverse power-law function is also reported to fit experimental data, though its use is less widespread than the KWW fit. \cite{Shimakawa1986, Schiff1995, Vardeny1980}
\begin{equation}
	\label{eq:Power_law_PPC_intro}
	f_{AD}(t) = f_{\infty} + \frac{A}{1+\left(t/\tau_0\right)^{\beta}}
\end{equation}
Both functions can be theoretically derived from a dispersive diffusion model, where the decay coefficient has a power-law decrease over time. The stretched exponential results from a so-called unimolecular process whereby the non-equilibrium sites undergo relaxation amongst a fixed density of mobile catalysts or ``walkers''.  On the other hand, an algebraic decay results from a so-called bimolecular process, whereby the non-equilibrium sites annihilate the mobile catalyst ``walkers'' upon relaxation. \cite{Ngai1981, Singh2003} Both the stretched exponential and algebraic decay represent limits of the same dispersive diffusion model with a tunable dispersion parameter. \cite{Singh2003}
As another option, one can envision the disorder-broadening of a single energy level into a Gaussian energy distribution, which also leads to a heavy-tail transient. \cite{Studenikin1998a, Marshall1987} 
All those models yield heavy-tail transients with fast initial responses and slow long-term responses, making it difficult to determine which model to apply for an arbitrary experimental transient. In addition, there is no single ``time constant'' for a heavy-tail transient, making it hard to determine when enough experimental data has been collected to accurately represent the time constant spectrum and predict an asymptotic value.  A model-independent analysis method is therefore needed.

This paper will introduce universal line-fits to characterize any heavy-tail experimental transient. The initial step reviewed in Section II is to plot the response in linear amplitude versus log-time where the transient represents a convolution of the time constant spectrum with an exponential decay. Here three different physically justifiable transient expressions are examined. For common physical systems, it is shown that the inflection point on such a plot represents the characteristic time scale of the transient.  Whereas previous efforts have attempted various deconvolution methods, these all suffer from the impracticality of collecting a complete transient dataset in log-time.  Instead, in Section III we conduct a lineshape analysis to find an empirical fit function with only four fit parameters which can extrapolate a finite dataset with sufficient accuracy.  It is shown that lineshapes must include a skewness parameter to distinguish various physically distinct time constant distributions.
The skew normal spectrum model and the dispersive diffusion transient model are introduced as simple four-parameter fits whereby one can estimate the peak position and asymptotic value of the time constant spectrum by identifying four features in a restricted dataset surrounding the inflection point in the original data. Unlike other analysis methods that require the entire duration of the response to be measured before the data is analyzed, \cite{Zorn2002, Gardner1959, Swingler1977} these four-parameter fits can accurately identify the key features from a truncated set of data, thereby saving orders of magnitude in measurement time.

In Section IV, analysis methods developed here are tested on published heavy-tail transient data from amorphous InGaZnO and amorphous silicon photoconductivity. 
These methods can also be applied to general heavy-tail relaxation problems.

\section{Theory and Modeling}
	\subsection{Decomposing heavy-tail transients into a time constant spectrum}
	Transient responses comprised of discrete time constants can be written as the sum of simple exponentials.
	For systems with a countable number of discrete time constants $\tau_1, \tau_2 \cdots \tau_n$, the overall response $f(t)$ over time $t$ has the form of Eq.~\eqref{eq:ME_discrete_form}, with amplitude $A_1, A_2 \cdots A_n$ for each time constant, and the asymptotic constant background ${f_{\infty}}$ at $t \to +\infty$.
	\begin{equation}
		\label{eq:ME_discrete_form}
		f(t)=f_{\infty} + \sum\limits_{i=1}^{n} {A_i{\mathrm{e}^{-t/\tau_i}}}
	\end{equation}
	
	\begin{figure*}
		\includegraphics[width=2\columnwidth]{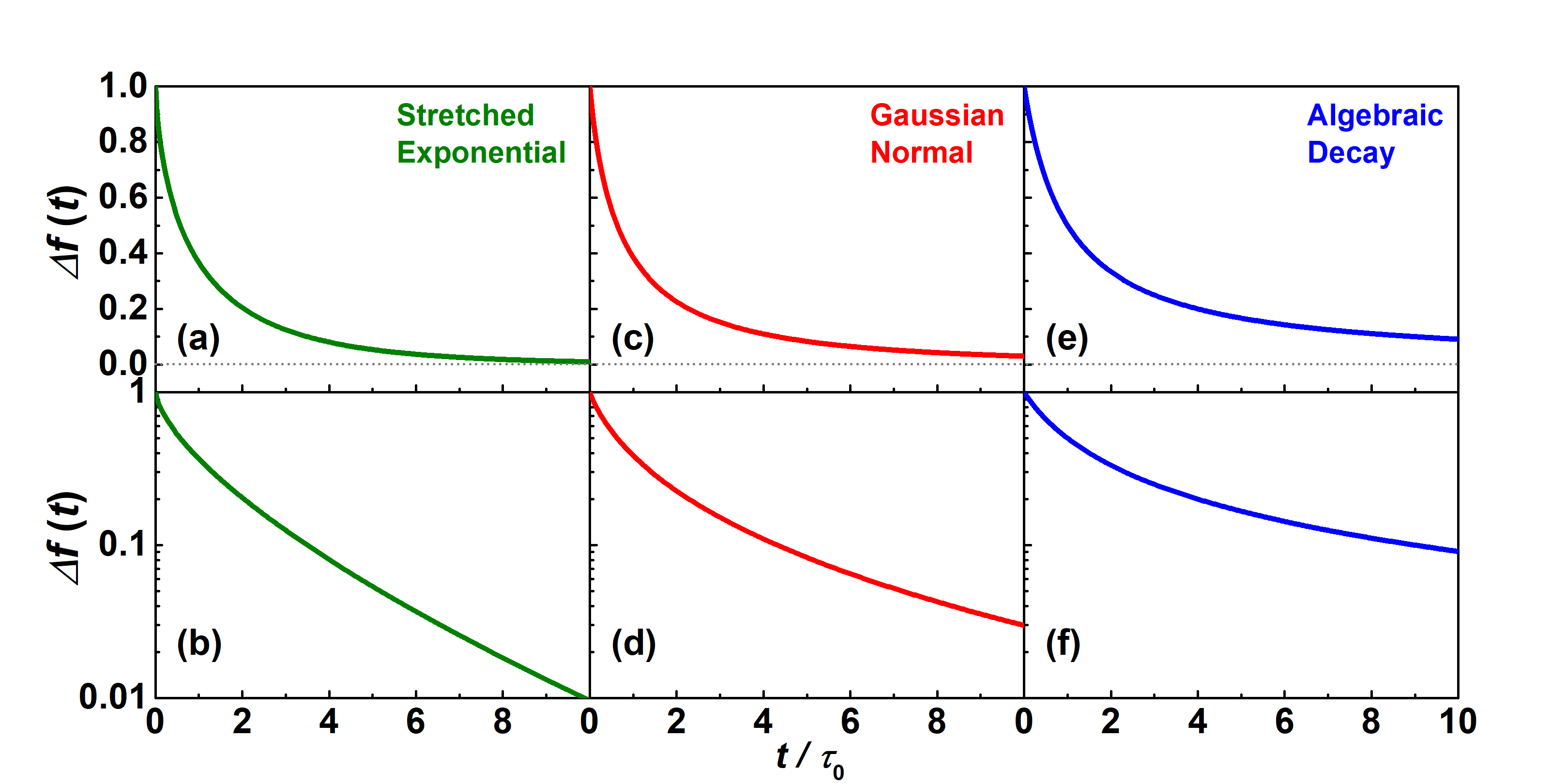}
		\caption{\label{fig:new_sync_lin} Plots of typical heavy-tail relaxations as a function of linear time. Panels (a, c, e) show the relaxations on linear-linear scale, and panels (b, d, f) show the same relaxations on log-linear scale. The stretched exponential decay (green) follows Eq.~\eqref{eq:SE_PPC}, Gaussian normal spectrum decay (red) follows Eq.~\eqref{eq:Gaussian_distri}, and algebraic decay (blue) follows Eq.~\eqref{eq:Power_law_PPC}.}
	\end{figure*}
	
	This equation can be generalized to a transient response with a continuum of amplitudes $g(\tau)\, d\tau$ defined for each time constant interval $d\tau$ around $\tau$.
	A generic transient response signal $f(t)$ can be expressed as a Fredholm integral equation of the first kind
	\begin{equation}
		\label{eq:general_integral}
		\begin{aligned}
			f (t) &= {f_{\infty}} + \int_{0}^{+\infty }{g (\tau){{\mathrm{e}}^{-t/\tau }} \, d\tau } \\
			&= {f_{\infty}} + \int_{0}^{+\infty }{g (\tau) h(t,\tau) \, d\tau }
		\end{aligned}
	\end{equation}
	where $g(\tau)$ is the time constant spectrum of interest; the simple exponential decay function $h(t,\tau)=e^{-t/\tau}$ is the kernel of the integral; and ${f_{\infty}}$ is the asymptotic constant background at $t \to +\infty$.
	
	This work will focus on systems with positive-definite $g(\tau)$. Such a condition is satisfied if and only if the transient response amplitude $f(t)$ decays monotonically with time, and if all higher derivatives also increase or decrease monotonically with alternating sign of derivative order. \cite{Zorn2002}
	\begin{equation}
	\label{eq:general_inequality}
	(-1)^n f^{(n)} (t) > 0, n = 1, 2, 3 \cdots
	\end{equation}
	In practice, this condition is satisfied for all transients provided there are no underdamped oscillations in the response.
	
	Several previous reports have tried to deduce the linear-scale time constant spectrum $g(\tau)$ directly from Eq.~\eqref{eq:general_integral}. 
	When the full transient $f(t)$ is known, $g(\tau)$ can be calculated through inverse Laplace transform. For example, Lindsey and Patterson have derived the time constant distribution $g(\tau)$ of stretched exponential relaxation in the form of an infinite series. \cite{Lindsey1980}
	But for an experimentally measured transient response $f(t)$ which also includes noise, deducing $g(\tau)$ is an ill-posed problem. For such an inversion of the Fredholm integral equation of the first kind, a small noise in the response $f(t)$ can lead to a large error in the deduced $g(\tau)$ distribution. \cite{Nagase1999}
	Methods like Tikhonov regularization can be used to suppress the noise. \cite{Nagase1999,Luo2015} However, those methods are computationally complex, and result in artificial spikes in the deduced $g(\tau)$ when the measurements end before reaching the asymptotic background $f_{\infty}$. \cite{Luo2016}
	
	The three different types of heavy-tail transients $f(t)$ considered in the following analysis are plotted versus linear time in Fig.~\ref{fig:new_sync_lin}.
	In the linear-linear plots of panels (a, c, e), all heavy-tail decays obey the rules of a positive-definite time constant spectrum per Eq.~\eqref{eq:general_inequality}; and in the log-linear plots of panels (b, d, f), all curves have upwards curvature with decreasing slope amplitude, indicating a monotonically increasing time constant.
	However, other than indicating the existence of a spectrum of time constants, such relaxation plots make it difficult to quantify the time constant distribution $g(\tau)$ or to even identify how much data is sufficient to extract $g(\tau)$.
	Instead of log-linear plots, we shall demonstrate the advantage of linear-log plots in identifying when sufficient data is collected to characterize the transient, and in curve-fitting finite datasets of such transients.
	
	\begin{figure}[]
		\includegraphics[width=0.9\columnwidth]{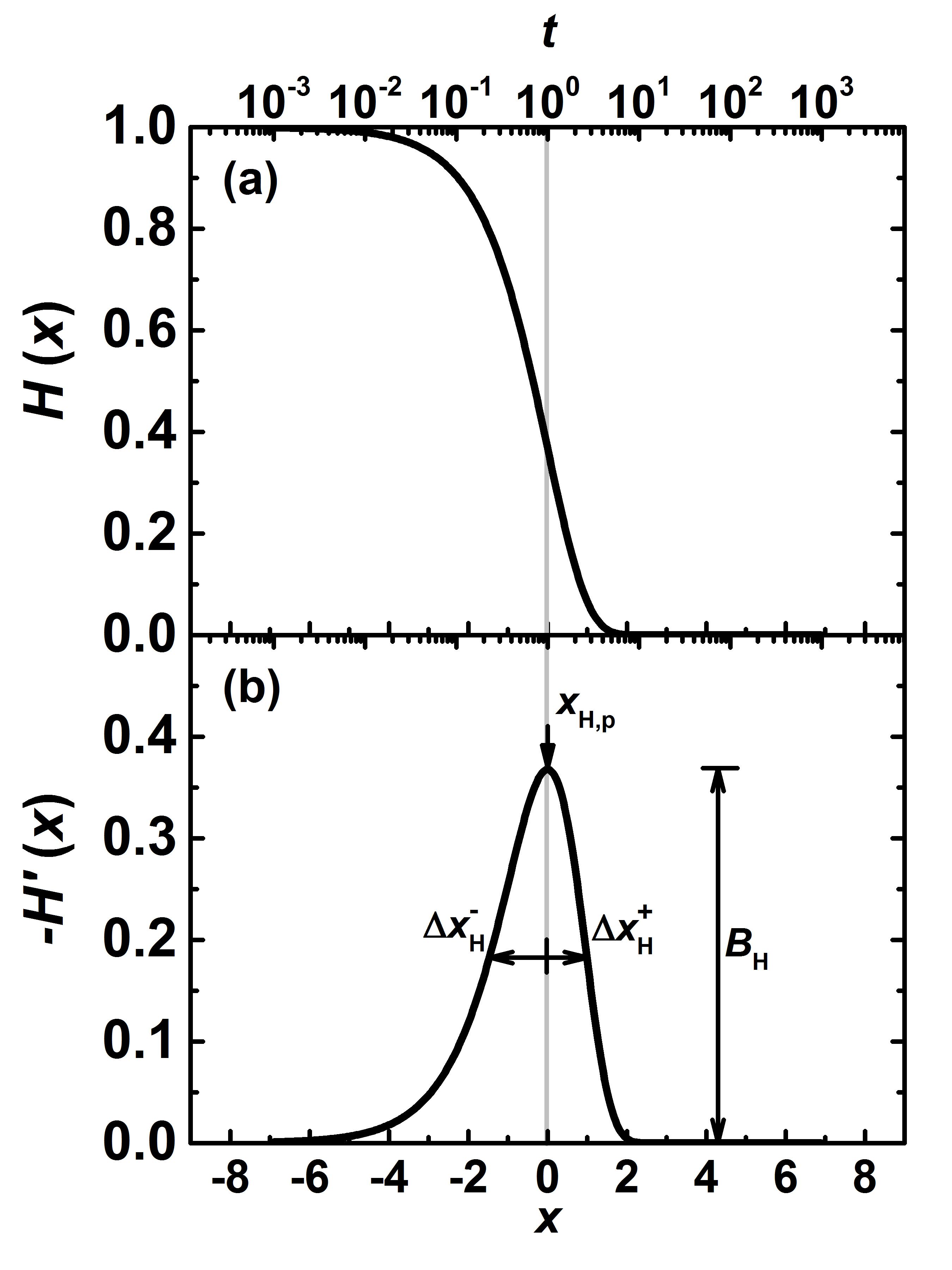}
		\caption{\label{fig:basic_struct} Plots of (a) the simple exponential decay impulse response $H(x)$, and (b) its derivative $-H'(x)$ with respect to log-time $x = \ln t$. The peak position of the derivative maximum $x_{H,p}$ is indicated with the downward arrow, and the peak height with $B_H$. The partial widths at half maximum $\Delta x^-_H$ and $\Delta x^+_H$ are indicated with the horizontal arrows.}
	\end{figure}
	
	As we shall see, many of these numerical difficulties can be circumvented by rewriting the transient response as a function of $\ln t$ instead of linear $t$, converting the inverse Fredholm integral into a far more tractable deconvolution problem.
	Following the first step of Gardner, \cite{Gardner1959} here we define variables $x = \ln t$ as the log-scale time and $u = \ln \tau$ as the log-scale time constant.
	Substituting $t = \mathrm{e}^x$ and $\tau = \mathrm{e}^u$ into Eq.~\eqref{eq:general_integral} gives the following expression for the log-scale transient $F(x)$, with log-scale time constant spectrum defined as $G(u)=\tau g(\tau)$:
	\begin{equation}
	\label{eq:log_integral}
	F(x)={f_{\infty}} + \int_{-\infty}^{+\infty }{G(u) {{\mathrm{e}}^{-\mathrm{e}^{x-u} }} \, du} .
	\end{equation}
	Note that $x$ appears only as a difference with $u$. If we define the impulse response $H(x)$
	\begin{equation}
	\label{eq:impulse}
	H(x) = \exp[-\exp(x)] ,
	\end{equation}
	the desired convolution form results:
	\begin{equation}
	\label{eq:convolution_form}
	F(x)={f_{\infty}} + \int_{-\infty}^{+\infty }{G(u) H(x-u) \, du} .
	\end{equation}
	The time constant spectrum $G(u)$ can be deduced through deconvolution of $F(x)$ using the Gardner transform \cite{Gardner1959} or other methods, and requires the full transient $F(x)$ to be known from $t = 0$ to $\infty$.
	The same time constant spectrum $G(u)$ can also be deduced from any order derivative of the log-scale transient response. For example, the first derivative of the log-scale transient is:
	\begin{equation}
		\label{eq:convolution_form_dev}
		F'(x)=\int_{-\infty}^{+\infty }{G(u) H'(x-u) \, du} .
	\end{equation} 
	Because the first derivative eliminates the need to fit the asymptotic offset value $f_{\infty}$, the method below will use the derivative $F'(x)$ to deduce $G(u)$.
	
	
	It is important to re-examine the familiar exponential decay in log-time, to develop an intuition concerning the deconvolution.  The exponential impulse response $H(x)$ in log-time is plotted in Fig.~\ref{fig:basic_struct}(a) and its derivative $H'(x)$ in Fig.~\ref{fig:basic_struct}(b). 
	The curve of $H(x)$ shows an inflection point, corresponding to the peak in $-H'(x)$ at the mode position (local maximum) $x_{H,p} = 0$, with mode amplitude $B_H = -H'(x_{H,p}) \approx 0.368$. Because $-H'(x)$ has unity integrated amplitude $\int_{-\infty}^{+\infty }-H'(x)dx=1$, it follows from Eq.~\eqref{eq:convolution_form_dev} that the integrated amplitude of the time constant spectrum $G(u)$ equals the total response amplitude.
	\begin{equation}
	\label{eq:integrated_amplitude}
	\begin{aligned}
	f(0) - f_{\infty}
	= -\int_{-\infty}^{+\infty} {F'(x) \, dx}
	= \int_{-\infty}^{+\infty} {G(u) \, du}
	\end{aligned}
	\end{equation}
	The impulse function $-H'(x)$ has a left-skewed unimodal (single-peak) shape, with half-maximum partial width $\Delta x_H^- \approx 1.46$ on the left and $\Delta x_H^+ \approx 0.99$ on the right. 
	
	Note that in log-time, every exponential has exactly the same linewidth. For a different value of the decay time constant, the curve merely translates in log-time.  Also, the linewidth of this curve represents the absolute minimum linewidth that any transient response can have in log-time. Any decay that is not a pure exponential will be a convolution of this curve with a time constant spectrum of finite width, and hence will be broadened with respect to this pure exponential.
	
	\subsection{Physical examples of heavy-tail time constant spectra}
	In this section we will discuss three physical examples of relaxation responses in terms of their time constant spectra $G(u)$.
	To establish an intuition for using $G(u)$ as the fingerprint for different physical mechanisms, the three models chosen here have distinct symmetries.
	The Gaussian normal distribution defines a symmetric $G(u)$ spectrum; the stretched exponential model yields a left-skewed $G(u)$ spectrum; and the algebraic decay model yields a right-skewed $G(u)$ spectrum.
	
	\begin{figure*}[]
		\includegraphics[width=2\columnwidth]{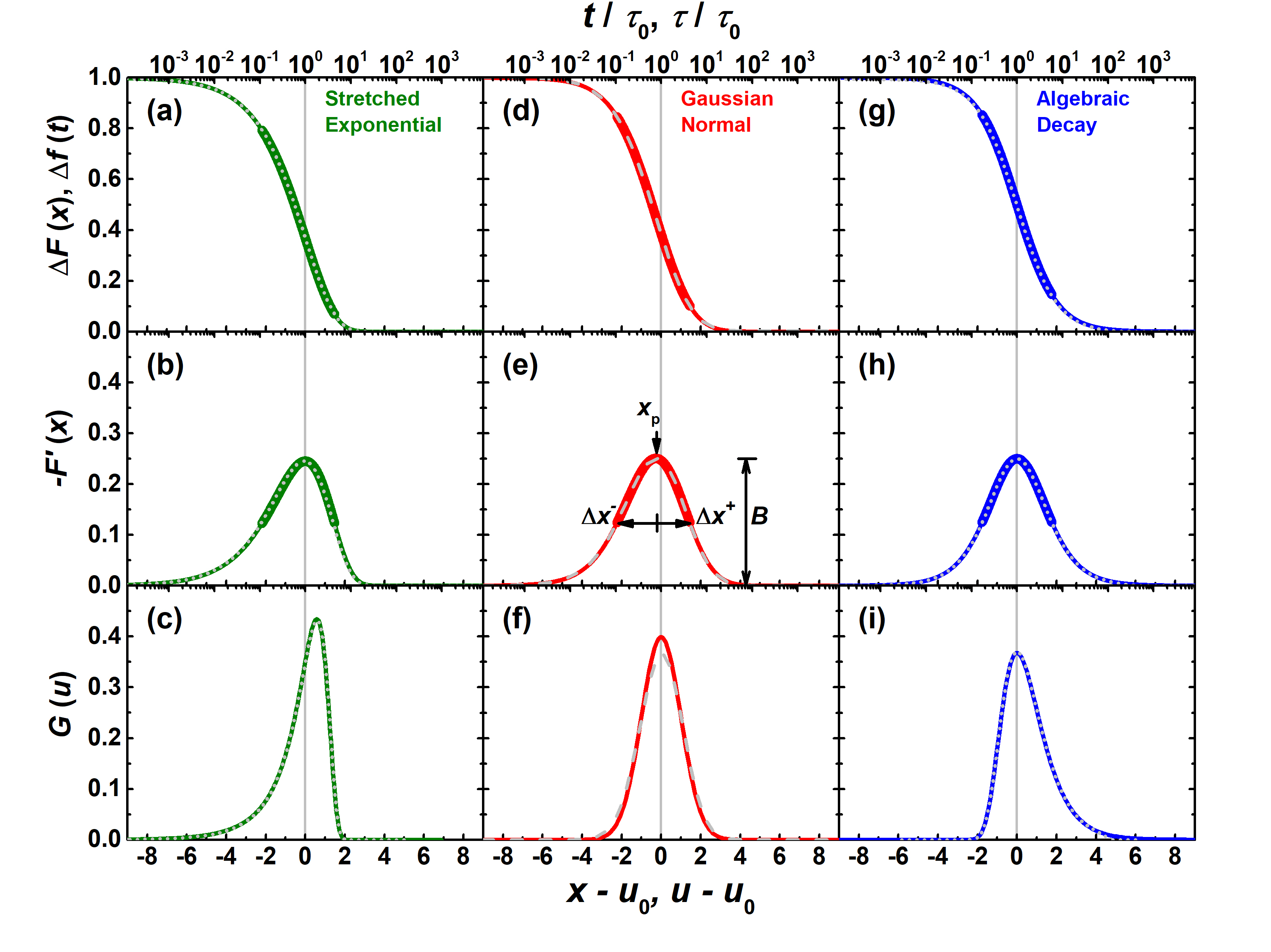}
		\caption{\label{fig:new_sync_data} Plots of model transients (green, red, blue) and their corresponding four-parameter fits from Section III (gray). The stretched exponential decay (a-c, green) follows Eq.~\eqref{eq:SE_PPC_log}, Gaussian normal spectrum (d-f, red) follows Eqs.~\eqref{eq:Gaussian_distri}, and algebraic decay (g-i, blue) follows Eq.~\eqref{eq:Power_law_PPC_log}. The top axis designates time $t$, and the bottom axis $x = \ln t$.
		Panels (a, d, g) show the decay responses $\Delta F(x)$ versus log-scale time. Panels (b, e, h) show the log-scale derivatives $-F'(x)$ of the transients. Panels (c, f, i) show the log-scale time constant spectra $G(u)$ of the transients.
		The gray dashed lines in the middle column (d-f) show the skew normal (SN) estimation corresponding to the Gaussian normal spectrum; and the gray dotted lines in the left (a-c) and right (g-i) columns show the dispersive diffusion (DD) estimation to the stretched exponential decay and the algebraic decay, respectively.
		Note the SN and DD fits are made only from the four curve parameters listed in Table \ref{tab:sync_data} from within the FWHM subsection in the derivative plots, drawn with thicker lines, yet are accurate to within a few percent, as per Fig.~\ref{fig:SN_error}.}
	\end{figure*}
	
	\begin{table*}[!hbtp]
		\caption{\label{tab:sync_data} Line shape features of the linear-log derivative plots in Fig.~\ref{fig:new_sync_data}(b, e, h) are listed in the left column.
		The SN fit parameters calculated from these features using Eqs.~\eqref{eq:empirical_equations} are listed in the middle column.
		Similarly, the DD fit parameters calculated using Eqs.~\eqref{eq:DD_empirical_equations} are listed in the right column.
		}
		\begin{ruledtabular}
			\begin{tabular}{c|cccc|cccc|cccc}
				\multirow{2}{*}{Model}
				& \multicolumn{4}{c|}{Derivative features} & \multicolumn{4}{c|}{SN fit parameters} & \multicolumn{4}{c}{DD fit parameters} \\
				\cline{2-13}
				& $x_p$   & $B$   & $\Delta x^-$   & $\Delta x^+$
				& $u_0$	  & $A$   & $s$   & $\alpha$
				& $u_0$   & $A$   & $\beta$   & $k$ \\ 
				\hline \hline
				Stretched exponential
				& 0    & 0.25   & 2.2   & 1.5 
				& 1.39    & 0.98   & 1.73   & -4 
				& 0.03    & 1.00   & 0.67   & 0.02 \\
				\hline
				Gaussian normal
				& -0.27    & 0.25   & 2.0   & 1.7
				& 0.66    & 1.00   & 1.29   & -0.88
				& 1.11    & 0.99   & 0.89   & 0.73	\\ 
				\hline 
				Algebraic decay
				& 0    & 0.25   & 1.8   & 1.8  
				& -0.48    & 0.96   & 1.37   & 1.6  
				& 4.60    & 0.99   & 1.00   & 0.99 
			\end{tabular}
		\end{ruledtabular}
	\end{table*}
	
	\subsubsection{Gaussian normal time constant spectrum}
	
	The physical model that most naturally describes a statistical distribution of independent time constants is the distributed activation energy model. \cite{Studenikin1998a, Marshall1987} Consider a thermally activated process with activation energy $E$, where each relaxation time constant $\tau$ is described with
	\begin{equation}
		\label{eq:E_to_tau}
		\tau =\frac{1}{\nu }{{\mathrm{e}}^{E/{{k}_{B}}T}}
	\end{equation}
	whereby $\nu$ is the attempt-to-escape frequency, often assumed to be phonon frequency, $k_B$ is Boltzmann’s constant, and $T$ is the temperature. \cite{Studenikin1998, Luo2013} 
	Thus the time constant $u$ in logarithmic time is linearly related to the activation energy $E$ through the following equation.
	\begin{equation}
		\label{eq:E_to_u}
		u = \ln \tau = E/k_BT - \ln \nu
	\end{equation}
	For systems with trap states spanning a distribution of activation energies $E$, the time constant spectrum $G(u)$ is therefore proportional to the density of trap states.
	
	A Gaussian normal time constant spectrum might naturally arise in a disordered system, where statistical disorder in the local configuration would lead to a homogeneously broadened Gaussian distribution of trap activation energies \cite{Tsormpatzoglou2014, Nagase1999} or equivalently, a Gaussian time constant spectrum in $\ln \tau$.
	The Gaussian is symmetric in log-time with zero skewness, and can be characterized by three parameters: its integrated spectral weight $A$, central value $u_0$, and standard deviation $\sigma$.
	\begin{equation}
	\label{eq:Gaussian_distri}
	G_{GN}(u)=\frac{A}{\sigma \sqrt{2\pi }} \mathrm{e}^{-\frac{( u-u_0 )^2}{2{\sigma}^2}}
	\end{equation}
	The corresponding transient response $F_{GN}(x)$ requires integrating from $-\infty$ to $+\infty$ per Eq.~\eqref{eq:convolution_form}. 
	
	An example of the Gaussian distribution model is plotted in the central column of Fig.~\ref{fig:new_sync_data} (red).
	The time constant spectrum $G_{GN}(u)$ with unit area is plotted in Fig.~\ref{fig:new_sync_data}(f), assuming a standard deviation $\sigma = 1.1$. The corresponding log-scale transient $F_{GN}(x)$, and log-scale derivative $-F_{GN}'(x)$ are plotted in panels (d) and (e), respectively.
	Note that because the Gaussian distribution model explicitly assumes a symmetric $G_{GN}(u)$, the log-scale derivative $-F'(x)$ is left-skewed, due to the convolution with the left-skewed impulse function $-H'(x)$.
	
	\subsubsection{Stretched exponential decay}
	The stretched exponential model assumes a specific functional form for the transient instead of assuming a time constant spectrum.
	The stretched exponential decay introduced in Section I is:
	\begin{equation}
		\label{eq:SE_PPC}
		f_{SE}(t) = f_{\infty} + A \mathrm{e}^{-(t/\tau_0)^{\beta}} .
	\end{equation}
	Again, the lineshape has three fit parameters plus an offset, where $A$ is again the total response amplitude, $\tau_0$ is the characteristic time scale of the stretched exponential, and $\beta$ is the stretching exponent.
	The corresponding log-time response is
	\begin{equation}
		\label{eq:SE_PPC_log}
		\begin{aligned}
			F_{SE}(x) &= f_{\infty} + A \cdot \mathrm{e}^{-\mathrm{e}^{\beta (x-u_0)}}\\
			&= f_{\infty} + A \cdot H[\beta (x-u_0)] ,
		\end{aligned}
	\end{equation}
	where the 2nd equation above clearly shows how the stretched exponent is scaled in log time by the stretching parameter $\beta$.
	
	This transient is predicted from the continuous-time random walk model of relaxation whereby mobile walkers that precipitate relaxation continue to migrate after catalyzing relaxation events at relaxation sites. \cite{Shlesinger1988, Shlesinger1984}
	Stretched exponential transients have been widely observed in many disordered, amorphous, and glassy systems. \cite{Phillips1996,Lee2008,Crandall1991,Luo2013,Kakalios1987,Bube1990,HossainChowdhury2013,Liu2008}
	
	At short time scales $t \to 0$, the initial transient behavior of the stretched exponential can be approximated as
	\begin{equation}
	\label{eq:SE_PPC_short_time}
	\begin{aligned}
	f_{SE}(x) & \cong f_{\infty} + A \left[ 1-(t/\tau_0)^{\beta} \right] ,\\
	F_{SE}(x) & \cong f_{\infty} + A \left[ 1-\mathrm{e}^{\beta (x-u_0)} \right] .\\
	\end{aligned}
	\end{equation}
	As we will see in the next section, this short time transient is identical to that of the algebraic decay.
	
	A stretched exponential decay can also be characterized with a rate equation that shows a power-law decay in time. The decay rate $f'(t)$ is proportional to the excess signal $f(t) - f_{\infty}$ by a time-dependent power law:
	\begin{equation}
	\label{eq:SE_decay_rate}
	\begin{aligned}
	f'(t) &\propto -t^{-(1-\beta)}[f(t)-f_{\infty}] .
	\end{aligned}
	\end{equation}
		 
	The stretched exponential satisfies the monotonic decay condition of Eq.~\eqref{eq:general_inequality}, and thus can be deconvolved to a positive-definite time constant spectrum $G_{SE}(u)$.
	Johnston has calculated the analytical function of $G_{SE}(u)$ for a few rational $\beta$ values, and has shown that $G_{SE}(u)$ always has a continuous left-skewed unimodal distribution. \cite{Johnston2006}
	For other $\beta$ values, $G_{SE}(u)$ can be numerically calculated through an inverse Laplace transform of Eq.~\eqref{eq:SE_PPC}. \cite{Hollenbeck1998}
	An example of a stretched exponential decay is plotted in the left column of Fig.~\ref{fig:new_sync_data} (green), with the stretching exponent $\beta = 2/3$ and unit response amplitude.
	Fig.~\ref{fig:new_sync_data}(a) plots the transient response versus log-scale time, and panels (b) and (c) plot its derivative $-F_{SE}'(x)$ and the time constant spectrum $G_{SE}(u)$, respectively. Both $-F_{SE}'(x)$ and $G_{SE}(u)$ are clearly left-skewed for the stretched exponential transient.
	
	\subsubsection{Algebraic decay}
	The algebraic decay is another kind of relaxation behavior experimentally reported, such as in the recombination of electron-hole pairs in amorphous silicon and amorphous chalcogenides. \cite{Shimakawa1986, Schiff1995, Vardeny1980} Similar to the stretched exponential relaxation, this form is also predicted from a variant of the continuous-time random walk model, but here the walkers are assumed to annihilate when they reach relaxation sites. \cite{Metzler2000, Ngai1981}
	Therefore, the decay rate $f'(t)$ is proportional to the \textit{square} of excess density $f(t) - f_{\infty}$, \cite{Shimakawa1986, Ngai1981} giving
	\begin{equation}
	\label{eq:Power_law_decay_rate}
	\begin{aligned}
	f'(t) &\propto -t^{-(1-\beta)}[f(t)-f_{\infty}]^2 .\\
	\end{aligned}
	\end{equation}
	This leads to the algebraic decay function
	\begin{equation}
	\label{eq:Power_law_PPC}
	f_{AD}(t) = f_{\infty} + \frac{A}{1+\left(t/\tau_0\right)^{\beta}}
	\end{equation}
	with three lineshape parameters and an offset, where $A$ is the total response amplitude, $\tau_0$ is the characteristic time scale, and $\beta$ is the dispersive coefficient.
	The corresponding log-time response is
	\begin{equation}
	\label{eq:Power_law_PPC_log}
	F_{AD}(x) = f_{\infty} + \frac{A}{1+\mathrm{e}^{\beta(x-u_0)}} .
	\end{equation}
	
	At short time scales $t \to 0$, the initial transient behavior of the algebraic decay can be approximated as
	\begin{equation}
	\label{eq:Power_law_short_time}
	\begin{aligned}
	f_{AD}(x) & \cong f_{\infty} + A \left[ 1-(t/\tau_0)^{\beta} \right] ,\\
	F_{AD}(x) & \cong f_{\infty} + A \left[ 1-\mathrm{e}^{\beta (x-u_0)} \right] ,\\
	\end{aligned}
	\end{equation}
	identical to Eq.~\eqref{eq:SE_PPC_short_time}. Thus an algebraic decay is indistinguishable from a stretched exponential decay at short time scales $t < \tau_0$. The algebraic decay will relax slower than the stretched exponential decay when the measurement duration is sufficiently long.
	
	At long times $t > \tau_0$, the asymptotic behavior describes an inverse power-law heavy-tail
	\begin{equation}
	\label{eq:Power_law_long time}
	f_{AD}(t) \cong f_{\infty} + A \left(t/\tau_0\right)^{-\beta},
	\end{equation}
	hence the algebraic decay is sometimes also referred to as an inverse power-law decay.
	
	The algebraic decay also satisfies the monotonic decay condition Eq.~\eqref{eq:general_inequality}, and can be deconvolved into a positive-definite time constant spectrum $G_{AD}(u)$. For the special case with $\beta = 1$, the analytical expression for $g_{AD}(\tau)$ and $G_{AD}(u)$ are
	\begin{equation}
	\label{eq:Power_law_beta_1}
	\begin{aligned}
	g_{AD,\beta=1}(\tau) &= \frac{\tau_0}{\tau^2}\cdot\mathrm{e}^{-\tau_0/\tau}\\
	G_{AD,\beta=1}(u) &= \mathrm{e}^{-(u-u_0)}\cdot H[-(u-u_0)]
	\end{aligned}
	\end{equation}
	For $0 < \beta < 1$, the time constant spectrum $G_{AD}(u)$ can be numerically deconvolved from Eq.~\eqref{eq:Power_law_PPC_log}.
	An algebraic decay with the $\beta = 1$ and unit response amplitude is plotted in Fig.~\ref{fig:new_sync_data}(g) (blue) versus log-scale time $x = \ln t$, with its log-scale derivative $-F_{AD}'(x)$ in panel (h) and log-scale time constant spectrum $G_{AD}(u)$ in panel (i).
	
	Note that an algebraic decay will always have a symmetric derivative in log-time $-F'_{AD}(x)$ around its central value $u_0$ as demonstrated in panel (k). After deconvolving with the left-skewed impulse function, the resulting time constant spectrum $G_{AD}(u)$ is always right-skewed as seen in panel (l).
	
	\section{Skewed four-parameter fits to time constant spectrum}
	In this section, we propose to fit the time constant spectrum with four parameters describing: amplitude, central value, linewidth, and, in particular, skewness. This is more versatile than any of the three-parameter models discussed above in Section II.B, whose skewness is constrained to be zero, positive, or negative, respectively. The skew normal (SN) spectrum lineshape and the dispersive diffusion (DD) transient function are introduced as examples of four-parameter models that include skewness. Both methods can fit reasonably well to any of the decay responses described above, with the DD model showing particularly excellent fits to the SE and AD models, and the SN showing excellent fits to the Gaussian.  Most importantly, these skewed four-parameter models allow extrapolation from a finite experimental dataset to an accurate asymptotic value.
	
	\subsection{Direct extraction of the time constant spectrum}
	As was mentioned in the introduction, several previous reports have attempted to extract $G(u)$ directly from a general decay response $F(x)$, relying on deconvolution. For example, by approximating the impulse function $-H'(x)$ in Eq.~\eqref{eq:convolution_form_dev} as a delta function, Jackson \textit{et al.} proposed to use the log-scale derivative $-F'(x)$ directly as a rough estimation of the time constant spectrum $G(u)$. \cite{Jackson1986} This delta-function approximation is only valid when the linewidth $\Delta x$ of the derivative peak $-F'(x)$ is significantly wider than the linewidth $\Delta x_H$ of the impulse function derivative $-H'(x)$, $\Delta x \gg \Delta x_H$, and such an estimated time constant spectrum can at best achieve a time resolution of the order $\Delta x_H$.
	
	To achieve a more accurate deconvolution of Eq.~\eqref{eq:convolution_form_dev} for narrower time constant spectra, the full Gardner transformation method can be used.  The Gardner transformation extracts $G(u)$ by Fourier transforming both sides of Eq.~\eqref{eq:convolution_form_dev}, thereby converting the deconvolution to simple division in the Fourier domain. \cite{Gardner1959} However, both the Fourier transform and the inverse Fourier transform steps require a complete dataset of the transient, integrating from $x = -\infty$ to $+\infty$.
	In real experiments, the measurement resolution may not be able to cover the initial response at arbitrarily short time scales ($x \to -\infty$), and the measurement duration can rarely cover the entire tail of the heavy-tail response at long times ($x \to +\infty$). Such a truncated dataset is known to cause artifacts of ``ripples'' in the Gardner transformed results, limiting its accuracy in analyzing experimental datasets. More recent improvements on the Gardner deconvolution have been proposed, but all suffer from the same limitations when handling a truncated experimental dataset. \cite{Jibia2012}
	
	Another way to extract the time constant spectrum $G(u)$ is through the moments of the log-scale derivative $-F'(x)$, as proposed by Zorn. \cite{Zorn2002}
	Using the properties of the convolution integral, the first, second, and third central moments of the time constant spectrum $G(u)$ can be calculated from those of the measured derivative $-F'(x)$. Then the best fitting model can be determined by comparing the calculated moments with those of all candidate models. However, similar to the Gardner transform, calculating the moments of $-F'(x)$ also requires integrating from $x = -\infty$ to $+\infty$, which induces a large error with a truncated dataset, making Zorn's method just as difficult to implement in practice.
	
	Finally, one can in principle deduce the time constant spectrum $G(u)$ from a numerical deconvolution of the transient using Eq.~\eqref{eq:convolution_form}, or from its derivative using Eq.~\eqref{eq:convolution_form_dev}.  However, again, the numerical  deconvolution task requires that a reasonable extrapolation curve be determined for a finite experimental dataset, and prior to the present work, no analysis provided a reasonable estimate for such an asymptotic curve. For this reason, we propose two empirical models that successfully approximate various different functional forms to fit a truncated dataset \textit{from only four lineshape parameters} and thereby serves as an accurate approximation to the full deconvolution problem.
	
	\subsection{Parametrization of transient lineshape parameters}
	Four key lineshape features in $-F'(x)$ are quantified to characterize an arbitrary heavy-tail transient, following the steps below:
	\begin{enumerate}
		\setcounter{enumi}{-1}
		\item Plot the measured transient in log-time $x = \ln t$, and plot the log-time derivative $-F'(x)$. Take data past the inflection point in $F(x)$ to the FWHM of the derivative $-F'(x)$.
		
		\item First identify the peak position $x_p$ in the derivative $-F'(x)$, which is also the inflection point of the original transient on the linear-log plot. 
		
		\item Find the amplitude of the derivative peak $B = -F'(x_p)$. 
		
		\item Identify the half maximum position $x^- < x_p$ so that $-F'(x^-) = B/2$, defining the left half maximum linewidth $\Delta x^- = x_p - x^-$.
		
		\item Identify the half maximum position $x^+ > x_p$ so that $-F'(x^+) = B/2$, defining the right half maximum linewidth $\Delta x^+ = x^+ - x_p$.
	\end{enumerate}
	These four lineshape features can be extracted from an incomplete dataset as long as the transient surrounding the FWHM of the derivative curve $-F'(x)$ is measured (thick lines in Fig.~\ref{fig:new_sync_data}). This achieves a significant advantage over deconvolution as well as the Fourier transform method of Gardner \cite{Gardner1959} and the moment method of Zorn, \cite{Zorn2002} which cannot handle truncated datasets, and requires remarkably little information about the original curve to achieve extremely accurate fits to the diverse functional forms studied here. The extracted features of the three theoretical models discussed in Section II are listed in Table \ref{tab:sync_data}.
	
	\begin{enumerate}
		\setcounter{enumi}{4}
		\item From the four lineshape features $x_p$, $B$, $\Delta x^-$, and $\Delta x^+$, deduce three intermediate parameters: FWHM $\Delta x$, asymmetry $\epsilon$, and fractional linewidth contribution $c$ whereby $0 < c < 1$.
		\begin{equation}
			\label{eq:intermediate_paras}
			\begin{aligned}
			&\Delta x = \Delta x^- + \Delta x^+ \\
			&\epsilon = (\Delta x^+ - \Delta x^-)/\Delta x \\
			&c = \sqrt{\frac{\Delta x^2 - \Delta x^2_H}{\Delta x^2}}
			\end{aligned}
		\end{equation}
	\end{enumerate}
	For a narrow peak with $c < 0.7$ ($\Delta x < 3.4$), the lineshape is determined mostly by the impulse function $-H'(x)$, regardless of $G(u)$ skewness, and the transient can be approximated by any of the three-parameter models. Conversely, for a wide peak with $c > 0.95$ ($\Delta x > 8$), the impulse function $-H'(x)$ can be approximated as a delta function, giving $G(u) \approx -F'(u)$, the limit discussed by Jackson \textit{et.~al.} \cite{Jackson1986} The skewed four-parameter models are therefore only necessary for intermediate width when $0.7 < c < 0.95$.
	
	\subsection{Generalization of the time constant spectrum: Skew normal distribution}
	Below we describe how a single-peak continuous time constant spectrum $G(u)$ characteristic of disordered systems can be approximated empirically with a skew normal distribution.
	All three of the previously considered lineshapes -- Gaussian normal, stretched exponential, and algebraic decay -- are distinguished by their skewness, so a generic lineshape with an additional skewness parameter should serve as a reasonable approximation to any of the above lineshapes, as we shall demonstrate. 
	
	The skew normal distribution function $G_{SN}(u)$ is defined with the four parameters: amplitude $A$, central time constant $u_0$, linewidth $s$, and skewness $\alpha$. \cite{Azzalini1985}
	\begin{equation}
	\label{eq:SN_fit_spectrum}
	\begin{aligned}
	G_{SN}(u) &= \frac{A}{\sqrt{2\pi}} \mathrm{e}^{-\frac{(u-u_0)^2}{2 s^2}} \left[ 1+\mathrm{erf} \left( \alpha\cdot\frac{u-u_0}{\sqrt{2}s} \right) \right] \\
	\end{aligned}
	\end{equation}
	The asymmetry parameter $\alpha$ can be any real number. For convenience, a bounded asymmetry parameter $-1 < \delta < 1$ is introduced that
	\begin{equation}
	\label{eq:empirical_b_asym}
	\delta = \alpha / \sqrt{1+\alpha^2} ,
	\end{equation}
	whereby $\alpha, \delta < 0$ represents a left-skewed time constant spectrum and $\alpha, \delta > 0$ represents right-skewed.
	
	The skew normal time constant spectrum can be directly constructed from the four derivative peak shape parameters $B$, $x_p$, $\Delta x$, and $\epsilon$:
	\begin{enumerate}[label=\arabic*a.]
		\setcounter{enumi}{5}
		\item Calculate the four skew normal parameters $A$, $u_0$, $s$ and $\delta$ ($\alpha$) using the following set of empirical equations.
		\begin{subequations}
			\label{eq:empirical_equations}
			\begin{align}
			\label{eq:empirical_a_asym}
			\delta &= \left[ \frac{\epsilon - ( 1 - c)\epsilon_H}{0.173 \cdot c} \right] ^{1/5} \\
			\label{eq:empirical_c_var}
			s &= 0.425 \cdot c \cdot \Delta x \cdot {\frac{1}{\sqrt{{1-2\delta^2 / \pi}}}} \\
			\label{eq:empirical_d_mode}
			u_0 &= x_p + \gamma \cdot c^2 - \sqrt{\frac{2}{\pi}}s \cdot \mathrm{erf} \left[ \delta\frac{\sqrt{\pi}}{2}\right] \\
			\label{eq:empirical_e_height}
			A &= B\cdot \Delta x / \left(0.9+0.04 \cdot c^2 \right)
			\end{align}
		\end{subequations}
	\end{enumerate}
	In Eq.~\eqref{eq:empirical_equations}, Euler's constant $\gamma \approx 0.5772$ is the distance of the mean position of impulse response $-H'(x)$ to the left of its peak, and $\epsilon_H \approx -0.192$ is the asymmetry of $-H'(x)$.
	
	The empirical equations for the four shape parameters in the skew normal spectrum are determined in the following way.
	Since $\epsilon$ is nearly independent of $\delta$ for small $|\delta|$, but increases dramatically as $|\delta| \to 1$, $\delta$ can be approximated by a high-integer root of $\epsilon$ in Eq.~\eqref{eq:empirical_a_asym}.
	The derivative peak width contributed by the time constant spectrum $c \cdot \Delta x$ is approximately proportional to the standard deviation of the skew normal $\sqrt{1-2\delta^2/\pi} \cdot s$, giving Eq.~\eqref{eq:empirical_c_var}.
	The shift of the derivative peak position $x_p$ from the skew-normal position parameter $u_0$ is due to both the asymmetry of the impulse function $-H'(x)$ and the spectrum asymmetry. The peak shift induced by the the asymmetric impulse function is approximated by the second term in Eq.~\eqref{eq:empirical_d_mode}. The asymmetry induced peak shift, equals the mean position of the skew-normal distribution $\sqrt{\frac{2}{\pi}} s \delta$ for small $\delta$ but deviates for large $\delta$, empirically approximated by the third term in Eq.~\eqref{eq:empirical_d_mode}.
	The integrated amplitude $A$ is approximated by the product of peak height $B$ and $\Delta x$, with the width dependent denominator in Eq.~\eqref{eq:empirical_e_height} as a correction for the heavy-tailed impulsed function.
	
	The skew normal time constant spectrum and the corresponding heavy-tail transient can be reconstructed from the four parameters.
	\begin{enumerate}[label=\arabic*a.]
		\setcounter{enumi}{6}
		\item Calculate the skew normal time constant spectrum $G_{SN}(u)$ by substituting the values of $A$, $s$, $u_0$, and $\alpha$ into Eq.~\eqref{eq:SN_fit_spectrum}.
		\item Calculate the log-scale transient $F(x)$ and thus the linear-scale transient $f(t)$ by substituting $G_{SN}(u)$ into Eq.~\eqref{eq:convolution_form}.
	\end{enumerate}
	
	\subsection{Generalization of the decay transient:  Dispersive diffusion model}
	Alternatively, one can empirically approximate heavy-tail transients directly with a four-parameter decay function. Here we use the dispersive diffusion transient function that unifies the stretched exponential decay and the algebraic decay forms. This functional form can be applied for any heavy-tail decay with derivative peak asymmetry $\epsilon_H \leqslant \epsilon \leqslant 0$.
	
	The dispersive diffusion decay function $f_{DD}(t)$ is defined with four parameters: amplitude $A$, characteristic time constant $\tau_0$ (or $u_0 = \ln \tau_0$), dispersiveness $\beta$, and skewness $k$. \cite{Singh2003,Luo2017}
	\begin{equation}
		\label{eq:DD_fit_transient}
		\begin{aligned}
		f_{DD}(t) =& \frac{A(1-k)}{\exp[{(t/\tau_0)^{\beta}}]-k} \\
		F_{DD}(x) =& \frac{A(1-k)}{H[\beta(x - u_0)]^{-1} - k}
		\end{aligned}
	\end{equation}
	Both the stretched exponential function and the algebraic decay function can be considered as limit cases of the dispersive diffusion function, with the skewness parameter $k$ in the dispersive diffusion function implicitly determines the skewness of the time constant spectrum. At the limit $k \to 0$, Eq.~\eqref{eq:DD_fit_transient} reduces to the stretched exponential function with a left skewed spectrum; and at the limit $k \to 1$, Eq.~\eqref{eq:DD_fit_transient} reduces to the algebraic decay function with a right skewed spectrum. Heavy-tail transients with intermediate spectrum skewness can be fitted with $0 < k < 1$.
	
	Similar to the skew normal approximation method, the parameters of the dispersive diffusion relaxation can also be directly estimated from the four derivative peak shape parameters $B$, $x_p$, $\Delta x$, and $\epsilon$:
	\begin{enumerate}[label=\arabic*b.]
		\setcounter{enumi}{5}
		\item Calculate the four dispersive diffusion parameters $A$, $u_0$, $\beta$ and $k$ using the following set of empirical equations.
		\begin{subequations}
			\label{eq:DD_empirical_equations}
			\begin{align}
			\label{eq:DD_empirical_a_asym}
			k &= \sum_{i=0}^{5} \kappa_i (10 \, \epsilon)^i \\
			\label{eq:DD_empirical_b_beta}
			\beta &= \left( \sum_{i=0}^{5} \delta_i k^i \right) / \Delta x \\
			\label{eq:DD_empirical_c_position}
			u_0 &= \sum_{i=0}^{3} m_i [\ln(1-k)]^i \\
			\label{eq:DD_empirical_d_amp}
			A &= B \cdot \Delta x / \left( \sum_{i=0}^{5} b_i k^i \right)
			\end{align}
		\end{subequations}
	\end{enumerate}
	Eqs.~\eqref{eq:DD_empirical_equations} use polynomial functions to empirically fit the inverse functions of the dispersive diffusion parameters with respect to the transient derivative lineshape parameters. The values of the polynomial parameters are listed in Table~\ref{tab:DD_poly} below.
	\begin{table}[!h]
		\centering
		\caption{Polynomial fit parameters in the dispersive diffusion estimation Eqs.~\eqref{eq:DD_empirical_equations}.}
		\label{tab:DD_poly}
		\begin{ruledtabular}
		\begin{tabular}{c|cccccc}
			$i$     & 0    & 1     & 2     & 3       & 4     & 5     \\ \hline
			\hline
			$\kappa_i$ & 1    & 1    & 2   & 2.43    & 1.33 & 0.28 \\ \hline
			$\delta_i$ & 2.45 & 0.67  & 1.29  & -3.33   & 5.55  & -3.09 \\ \hline
			$m_i$     & 0    & 0.80  & -0.12 & -0.02 & N/A   & N/A   \\ \hline
			$b_i$     & 0.90 & 0.02 & 0.19  & -0.71   & 1.11  & -0.64
		\end{tabular}
		\end{ruledtabular}
	\end{table}
	
	With the dispersive diffusion parameters, one can reconstruct the entire heavy-tail transient response and its time constant spectrum.
	\begin{enumerate}[label=\arabic*b.]
		\setcounter{enumi}{6}
		\item Calculate the dispersive diffusion decay transient $f_{DD}{(t)}$ and $F_{DD}(x)$ by substituting the values of $A$, $u_0$, $\beta$, and $k$ into Eq.~\eqref{eq:DD_fit_transient}.
		\item Calculate the linear-scale time constant distribution $g(\tau)$ by numerical inverse Laplace transform of $f_{DD}{(t)}$, or the log-scale time constant spectrum $G(u)$ by numerical deconvolution of $F_{DD}(x)$.
	\end{enumerate}
	
	\subsection{Evaluation of fit accuracy}
	\begin{figure}[]
		\includegraphics[width=\columnwidth]{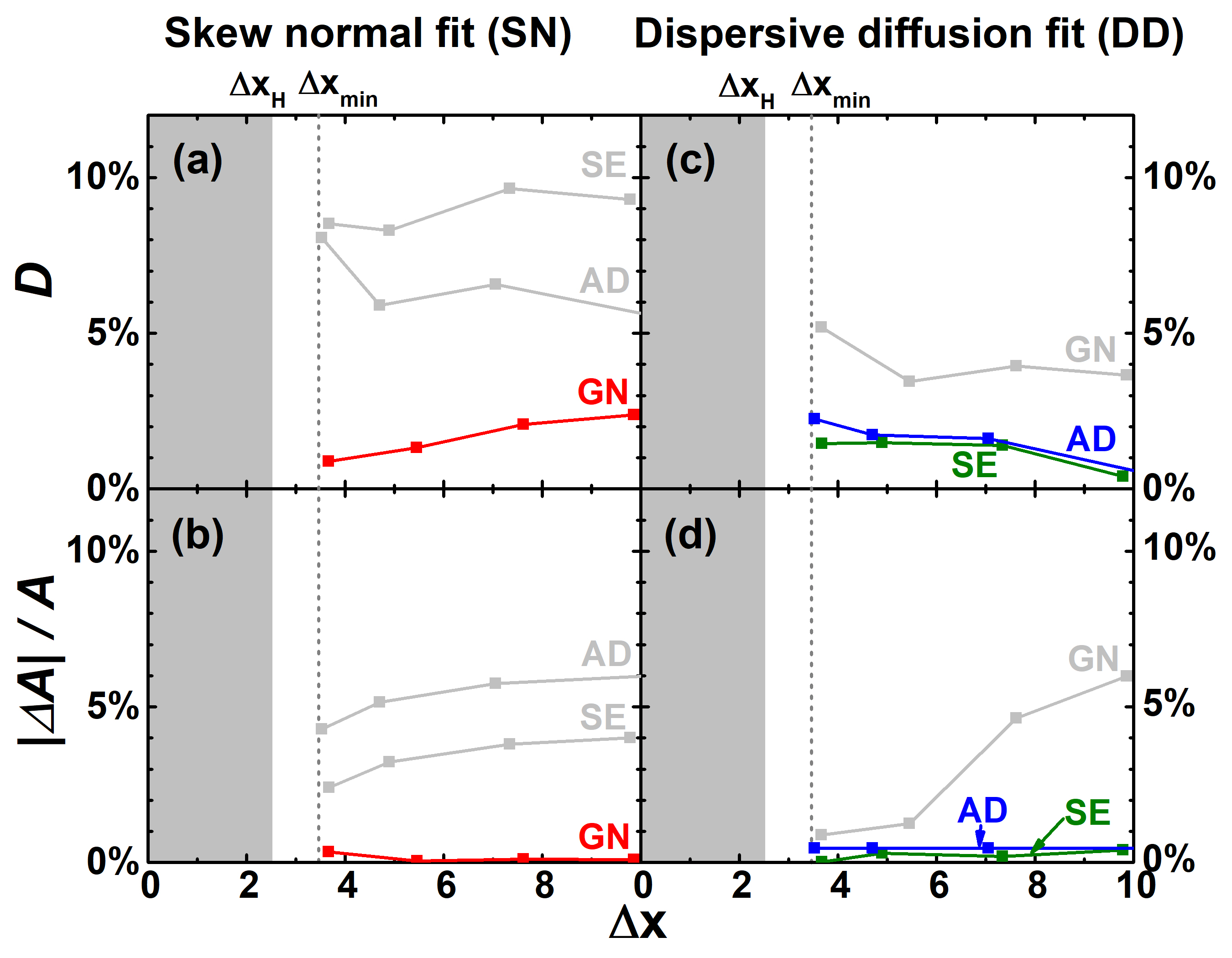}
		\caption{\label{fig:SN_error} Spectrum difference $D$ (a, c) and asymptotic amplitude error $|\Delta A| / A$ (b, d) between the skewed four-parameter approximation and the modeled time constant spectra as a function of spectral linewidth $\Delta x$ for various Gaussian normal spectrum transients (GN), stretched exponential transients (SE), and algebraic decay transients (AD).
		The left column show the results using the skew normal (SN) method of Section III.C and the right column for the dispersive diffusion (DD) method of Section III.D. 
		The vertical dashed line at $\Delta x_{\min} \approx 3.4$ indicates the minimum derivative linewidth needed to apply four-parameter fit methods. Heavy-tail transients cannot possess a linewidth narrower than $\Delta x = \Delta x_H$, indicated in gray.}
	\end{figure}
	
	Both the skew normal (SN) and the dispersive diffusion (DD) methods have been tested on the three theoretical heavy-tail transients discussed in Section II. The four lineshape features according to Section III.B, the resulting skew normal fit parameters according to Section III.C, and dispersive diffusion fit parameters according to Section III.D for the three models are listed in Table \ref{tab:sync_data}. Fig.~\ref{fig:new_sync_data} shows how well the skew normal time constant approximation (gray dashed lines) and the dispersive diffusion transient approximation (gray dotted lines) can match the various mathematical models (solid green, red, and blue lines). Both methods are able to reproduce the lineshape within the FWHM region of the $F'(x)$ derivative almost perfectly, and the error is mostly in the tails of the derivatives.
	
	To estimate the time constant spectrum accuracy of the approximation methods, the difference between the fit spectrum $G_{\mathrm{fit}}(u)$ and the original spectrum $G(u)$ is quantified in terms of the fraction of the non-overlapping area:
	\begin{equation}
	\label{eq:SN_fit_error}
	D  = 1 - \Lambda = 1 - \frac{\int_{-\infty}^{+\infty} {\min\left[G(u), G_{\mathrm{fit}}(u)\right] \, du}} {\int_{-\infty}^{+\infty} {\frac{1}{2} \left[G(u) + G_{\mathrm{fit}}(u)\right] \, du}} ,
	\end{equation}
	where $\Lambda$ is defined analogous to the overlapping coefficient of probability distributions, \cite{Inman1989} but normalized by the average of the original and the fit spectra.
	Fig.~\ref{fig:SN_error}(a,c) shows how $D$ changes with the derivative linewidth $\Delta x$, reproducing the time constant spectra within less than 10 \% error for the SN method and less than 5 \% error for the DD method.  But if the underlying physical mechanism is known, such that the SN fit is applied to Gaussian transients and the DD fit to SE and AD transients, this error is reduced below 3 \%.
	In Fig.~\ref{fig:SN_error}(b,d), the amplitude error $|\Delta A| / A$, defined as $\Delta A = A - A_{\mathrm{fit}}$ is equivalent to the accuracy of the asymptotic value.  This amplitude error is less than 6 \% for all three physical models using either approximation method.
	Again, if the functional form is known, then the SN to Gaussian transients and the DD fit to SE and AD transients are accurate to less than 1 \%.
	This level of accuracy allows one to extrapolate the entire heavy-tail transient to extremely long time scales which might exceed practical measurement duration, while helping to identify the underlying physical model.
	
 \section{Application to Experimental heavy-tail transients}
	\begin{figure*}[]
		\includegraphics[width=2\columnwidth]{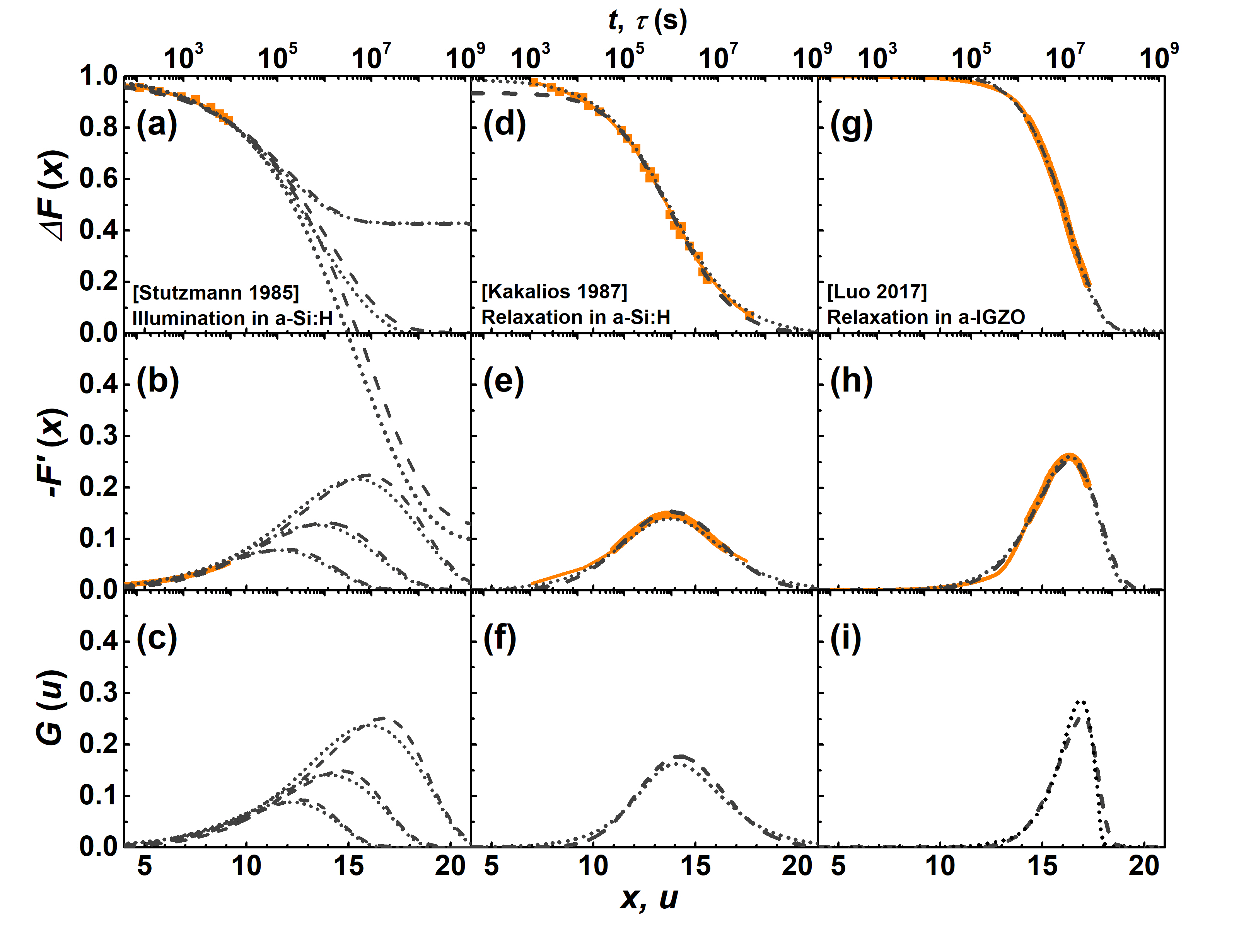}
		\caption{\label{fig:data_fit} 
			(a, d, g) Normalized transient responses $\Delta F(x)$ from three samples in literature, plotted versus log-time. (b, e, h) Derivative of the normalized log-time transient response. (c, f, i) Normalized time constant spectra ${G}(u)$ estimated from the transient responses. The orange squares are the discrete data points; and the orange lines are the continuous lines generated by smoothing discrete data points.
			The skew normal approximations are shown as the gray dashed lines; and the dispersive diffusion approximations are shown as the gray dotted lines.
			The heavy-tail transients in panels (a, d, g) are extracted from references \onlinecite{Stutzmann1985}, \onlinecite{Kakalios1987}, and \onlinecite{Luo2017}, respectively.}
	\end{figure*}
	
	\begin{table*}[!hbtp]
		\caption{\label{tab:fit_data} Line shape characteristics of the linear-log derivative peaks for datasets from the literature.
		The SN fit parameters are calculated from Eqs.~\eqref{eq:empirical_equations}; and the DD fit parameters from Eqs.~\eqref{eq:DD_empirical_equations}.
		}
		\begin{ruledtabular}
			\begin{tabular}{l|cccc|cccc|cccc}
				\multirow{2}{*}{Sample}
				& \multicolumn{4}{c|}{Derivative features} & \multicolumn{4}{c|}{SN fit parameters} & \multicolumn{4}{c}{DD fit parameters} \\
				\cline{2-13}
				& $x_p$   & $B$   & $\Delta x^-$   & $\Delta x^+$
				& $u_0$	  & $A$   & $s$   & $\alpha$
				& $u_0$   & $A$   & $\beta$   & $k$ \\ 
				\hline 
				\hline
				\multirow{3}{*}{\#1 a-Si:H}
				& 11.42    & 0.078   & 3.83   & 2.85   
				& 14.46    & 0.56   & 4.12   & -3.5 
				& 12.56    & 0.57   & 0.43   & 0.43 	\\
				
				& 13.36    & 0.13   & 4.19   & 3.12   
				& 16.69    & 0.90   & 4.58   & -3.6 
				& 14.60    & 1.02   & 0.39   & 0.43 	\\
				
				& 15.33    & 0.22   & 4.39   & 3.27   
				& 18.82    & 1.77   & 4.82   & -3.6 
				& 16.63    & 1.81   & 0.37   & 0.43		\\
				\hline
				\#2 a-Si:H
				& 13.75    & 0.15   & 2.93   & 2.97
				& 12.97    & 0.94   & 2.57   & 0.95
				& 21.46    & 0.99   & 0.60   & 0.99		\\
				\hline
				\#3 a-IGZO
				& 16.32    & 0.26   & 2.09   & 1.48   
				& 17.71    & 1.01   & 1.74   & -4 
				& 16.69    & 1.02   & 0.75   & 0.28 	
			\end{tabular}
		\end{ruledtabular}
	\end{table*}
	
	Finally, we apply the methods of Section III to experimental data from the literature, and analyze their time constant spectra with these skewed four-parameter models. The datasets of Samples \#1 and \#2 are extracted from reports on amorphous silicon (a-Si:H) and Sample \#3 from amorphous InGaZnO (a-IGZO). 
	Sample \#1 is a-Si:H from Stutzmann \textit{et.~al.}, \cite{Stutzmann1985} whereby the transient is proposed by the authors to result from an increase of the number of dangling bonds under constant illumination for 5 hours.
	Sample \#2 is a-Si:H from Kakalios \textit{et.~al.}, \cite{Kakalios1987} where room temperature defect density relaxation induced by thermal stress was measured for over a year.
	It was reported that defect creation/annihilation kinetics in a-Si:H is dominated by the diffusion of hydrogen passivation atoms,\cite{Kakalios1987, Morigaki1988} thus a stretched-exponential behavior was expected. \cite{Redfield1989, Bube1989a}
	Sample \#3 is an a-IGZO thin film from Luo \textit{et.~al.} \cite{Luo2017} The room temperature photoconductivity relaxation after saturation illumination was measured for around one year.
	Both stretched exponential and algebraic decay models are physically justifiable candidates for the heavy-tail transients observed in previous a-IGZO studies, \cite{Luo2013, Jeong2013, Han2014, Hsieh2014} so the DD fit is expected to perform better.
	
	To make a clear comparison among all datasets, each transient is normalized to a unit transient with zero asymptote.
	Sample \#1 has a discrete dataset shown in Fig.~\ref{fig:data_fit}(a) with its derivative plotted in panel (b).  This is an example of an insufficient dataset where the data was not taken past the inflection point in log-time.  The three pairs of fits match the measured segment of data equally well, and in panels \ref{fig:data_fit}(b) and (c) illustrate what can and what cannot be discerned when the measurement duration is not long enough.
	All fit curves overlap the measured transient and its derivative, but they imply very different time constant spectra in panel (c), and very different asymptotic values in panel (a). Thus one must always measure up to the inflection point on the linear-log plot of the transient, or equivalently, to the peak on the log-scale derivative, to have a reasonable estimate of the characteristic time scale; and up to the right half-maximum point to be able to estimate the full lineshape and predict the asymptotic value.
	The original report described the transient with a power-law increase with illumination time $t$. As Redfield and Bube have pointed out later, \cite{Redfield1989} this power-law relationship at short times can alternatively be explained as a segment of a stretched exponential response, and we argue here that it could also just as easily be an algebraic response.	
	
	Sample \#2 has a discrete dataset shown in Fig.~\ref{fig:data_fit}(d), with its interpolated derivative plotted in panel (e). The derivative shows a clear peak as well as the half maximum point on both left and right sides.
	The lineshape parameters of the experimental transient and the corresponding four-parameter fits are listed in Table \ref{tab:fit_data}.
	With just the information of the peak and the two half maximum points, both the skew normal approximation and the dispersive diffusion approximation are able to accurately reconstruct the derivative and the transient in the FWHM region with time scales spanning 3 orders of magnitude. Whereas the skew normal function slightly underestimates the short-time tail of the time constant spectrum, causing 6 \% error in the estimated response amplitude, the dispersive diffusion fit, as expected, matches the experimental data with high accuracy. Note that even though the original report fitted the transient with the stretched exponential function, the symmetric derivative peak $\Delta x^- \simeq \Delta x^+$ suggests that the algebraic decay model might make a better three-parameter fit, as per the discussion at the end of Section II.B.3.
	
	Sample \#3 has a continuous dataset shown in Fig.~\ref{fig:data_fit}(g), with its derivative plotted in panel (h).  The derivative shows a clear peak as well as the left half maximum point. The right half maximum point is easily extrapolated from the trend in experimental data. The asymmetric derivative peak $\Delta x^- > \Delta x^+$ suggests that the dispersive diffusion fit might tend more towards a stretched exponential $k = 0$ in Eq.~\eqref{eq:DD_fit_transient}, and away from an algebraic decay $k = 1$.  Sure enough, the values listed in Table~\ref{tab:fit_data} reveal $k = 0.28$.
	The skew normal approximation and the dispersive diffusion approximation, as plotted by the gray dashed and dotted lines respectively, provide good fits to the transient and the derivative peak around the inflection point spanning time scales over 2 orders of magnitude, with error arising only at the tail of the derivative peak.
	These skewed four-parameter models will allow accurate prediction of the asymptotic value of the transient, which would otherwise take another ten years to measure experimentally.
	
	As shown in the above examples, the best way to represent a heavy-tail transient is on a linear-log plot that plots data as linear amplitude vs. log-time. Many previous reports showed the transient data on linear plots \cite{Stutzmann1985, Lee2010, Chen2010a, Luo2013}, log-log plots \cite{Stutzmann1985, Lee2010}, or log-linear plots with amplitude on log-scale and $t$ on linear scale \cite{Ghaffarzadeh2010, Steenbergen2011, Connelly2016}. Those representations of the data make it difficult to identify the important features of the transient lineshape, which can inadvertently lead to insufficient data collection and/or misinterpretation of the transient behavior.

\section{Conclusion}
	In conclusion, we analyzed heavy-tail transients described by various characteristic lineshapes as well as published experimental data in terms to evaluate their time constant spectra.
	By plotting a transient response on the linear-log plot, a single peak arises in the log-scale derivative for most experimentally measured heavy-tail responses. This derivative peak was shown to be a convolution of the log-scale time constant spectrum with an impulse function $-H'(x)$ of an exponential in log-time.
	During an experiment, it becomes possible to identify whether the measurement duration is long enough to extract the time constant spectrum by monitoring the derivative of the transient data in log-time. With a truncated dataset spanning the full-width at half maximum of the log-scale derivative, sufficient lineshape information is acquired to be able to reconstruct the entire time constant spectrum with a skewed four parameter model, as well as accurately estimate the asymptotic value of the transient decay.

\begin{acknowledgments}
This work was supported by the NSF MRSEC program DMR-1121262 at the Materials Research Center of Northwestern University and AFOSR Grant FA9550-15-1-0247. JL is grateful for formative discussions with Stephan Kim, and MG is grateful for discussions with Michael F. Shlesinger, James Kakalios, Elizabeth Steenbergen, Patrick Kung, John Torkelson, and Mizhou Wang.
\end{acknowledgments}

\bibliography{bibfile}

\end{document}